\documentclass[pra,twocolumn,showpacs,amsmath,amssymb]{revtex4}

\usepackage{graphicx}
\usepackage{dcolumn}
\usepackage{bm}
\usepackage{amssymb}
\usepackage{amsmath}
\usepackage{epsfig}

\begin{document}

\title{Decoherence effects on the quantum spin channels}
\author{Jian-Ming Cai}
\author{Zheng-Wei Zhou}
\email{zwzhou@ustc.edu.cn}
\author{Guang-Can Guo}
\affiliation{Key Laboratory of Quantum Information, University of
Science and Technology of China, Chinese Academy of Sciences,
Hefei, Anhui 230026, China}

\begin{abstract}
An open ended spin chain can serves as a quantum data bus for the
coherent transfer of quantum state information. In this paper, we
investigate the efficiency of such quantum spin channels which work
in a decoherence environment. Our results show that, the decoherence
will significantly reduce the fidelity of quantum communication
through the spin channels. Generally speaking, as the distance
increases, the decoherence effects become more serious, which will
put some constraints on the spin chains for long distance quantum
state transfer.
\end{abstract}

\pacs{03.67.Hk, 05.50.+q, 32.80.Lg}

\maketitle
\section{Introduction}
Quantum computation has the potential to outperform their
classical counterparts in solving some intractable problems which
would need exponentially longer time for a classical computer
\cite{Chuang,Nielsen}. Lots of efforts have been devoted to
searching for various kinds of real physical systems that maybe
appropriate for the implementation of quantum computation. One key
feature of such physical systems is scalability \cite{DiVincenzo}.
There are several prospective candidates for scalable quantum
computation, such as optical lattices \cite{Optical
Lattice1,Optical Lattice2,Optical Lattice3}, arrays of quantum
dots \cite{QDot1,QDot2,QDot3}, superconducting circuits
\cite{Superconducting1,Superconducting2}. It is also known that
universal quantum computation can be performed by a chain of qubit
with nearest neighbor Heisenberg or XY coupling together with some
other physical resources \cite{QC1,QC2,QC3,QC4}.

In the large-scale quantum computing, how to transmitting quantum
states from one location to the other location, which is a little
similar but not all the same to quantum information distributing
\cite{QID}, is an important problem. The primitive scheme of
quantum communication through an unmodulated spin chain is
proposed by S. Bose \cite{Bose}. It was shown that, quantum states
can be transferred via an open ended spin chain with ferromagnetic
Heisenberg interactions. The fidelity will exceed the highest
fidelity for classical transmission of a quantum state until the
chain length $N$ is larger than $80$. In Refs \cite{Christandl}.,
M. Christandal \emph{et al}. put forward a special class of
Hamiltonian that is mirror-periodic. Based on a spin chain with
such a mirror-periodic Hamiltonian, perfect quantum state transfer
can be achieved. Up to now, there are many other variational
schemes for the transfer of quantum states in spin systems
\cite{QWire,Burgarth}. In the real physical systems, especially
for solid state system, decoherence and noise is inevitable
\cite{Chuang}. For example, in the system of arrays of quantum
dots, both the surrounding nuclei spin environment \cite{SE} and
$1/f$ noise will induce decoherence. Therefore, under the
influence of decoherence, how efficient different spin chain
channels will work becomes an interesting and important problem.
On the other hand, dynamical decoherence properties of many-body
systems \cite{DSolid,Ziman,Khveshchenko} are basically significant
by itself. There have been several work about the decoherence and
spin chain channels \cite{Giovannetti,Daniel,Key}. However, the
decoherence effects on the efficiency of these quantum spin
channels have not yet been \textit{thoroughly} investigated.

In this paper, we calculate the fidelity of quantum communication
through the spin chain channels under the influence of decoherence.
Two representative kinds of environment model are investigated. One
is the one common spin environment \cite{Zurek}. The other is the
local independent environment \cite{IDE}. We show that the
efficiency of the spin channels will be significantly lowered by the
decoherence environment. As the spin chain length increases, the
decoherence effects may become very severe, which suggest some new
constraints on the spin chains for long distance quantum state
transfer. We mostly concentrate on the Heisenberg spin chain and the
mirror-periodic Hamiltonian scheme. However, some of the results are
applicable for other schemes of quantum state transfer.

The structure of this paper is as follows. In Sec. II we investigate
the efficiency of quantum spin channels in one common spin
environment. In Sec. III the situation of local independent
environment is discussed. In Sec. IV are conclusions and some
discussions.

\section{One common spin environment}
We start by considering the important decoherence model in spin
systems, i.e. one common spin environment. The Hamiltonian of the
spin chain with $N$ spins is denoted as $H_{S}$. The total
z-component of the spin is conserved, i.e.
$[\sum_{i=1}^{N}\sigma_{i}^{z},H_{S}]=0$. This is true for several
important spin chain channels \cite{Bose,Christandl}. The central
system interacts with one common spin environment $\Xi$
\cite{Zurek}, which is formed by $M$ independent spins, for large
values of $M$, as depicted in Fig. 1.
\begin{equation}
H_{S\Xi}=\frac{1}{2}\sum\limits_{i=1}^{N}\sigma_{i}^{z}\otimes
\sum\limits_{k=1}^{M}g_{k}\sigma_{k}^{z}
\end{equation}

\begin{figure}[htb]
\epsfig{file=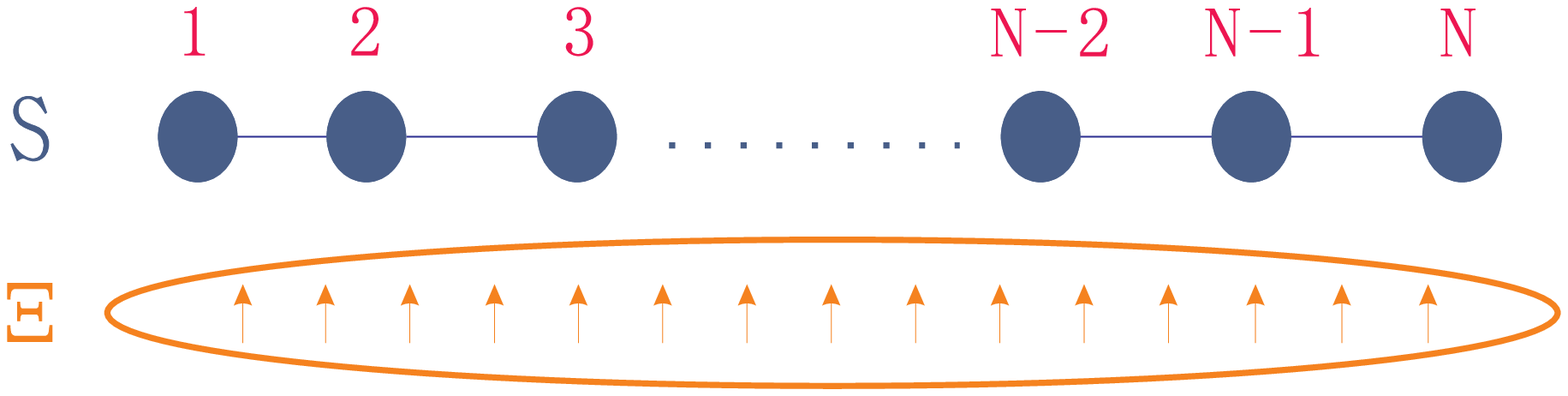,width=7cm} \caption{(Color online) A quantum
wire with $N$ spins in the line coupled with one common spin
environment.}
\end{figure}

The whole system of the spin chain $S$ and the environment $\Xi$ is
described by the Hamiltonian
\begin{equation}
H_{\mathcal{T}}=H_{S}+H_{S\Xi}
\end{equation}
Here the self-Hamiltonian of the environment $\Xi$ is neglected.
This simple decoherence model is an important solvable model of
decoherence, which are much relevant to quantum information
processing \cite{Chuang}. We will demonstrate how efficient the
quantum spin chain channels will work in such a decoherence
environment.

The quantum state to be transferred is located at the $1st$ spin,
$|\varphi_{in}\rangle=\alpha|0\rangle+\beta|1\rangle$. The initial
state of the central system is
\begin{equation}
|\psi_{S}(0)\rangle=\alpha|\mathbf{0}\rangle+\beta|\mathbf{1}\rangle
\end{equation}
with $|\mathbf{0}\rangle=|00\cdots 0\rangle$,
$|\mathbf{1}\rangle=|10\cdots 0\rangle$. In the following of this
paper, we will denote $|\mathbf{j}\rangle$ the state in which the
$jth$ spin is in the $|1\rangle$ state, while all the other spins
are in the $|0\rangle$ state. Without the influence of decoherence,
the transfer fidelity at time $t$ is defined as
$f_{1N}(t):=\langle\mathbf{N}|e^{-iH_{s}t}|\mathbf{1}\rangle$. Using
the calculations in \cite{Christandl2}, we can write down the final
density matrix $\rho_N(t)$ in the absence of the decoherence, just
in terms of the initial coefficients of the input state, and the
transfer fidelity, as follows
\begin{equation}
\rho_{N}(t)=\left(
\begin{array}{cccc}
1-|f_{1N}(t)|^2 & \alpha f_{1N}^{*}(t)\\
\alpha^{*}f_{1N}(t) & |f_{1N}(t)|^2
\end{array} \right)
\end{equation}

Now we take into account the interaction between the spin chain and
the environment. We denote the basis of the environment $\{|m\rangle
\langle m|\}$, $\sum_{m=0}^{2^M-1}|m\rangle \langle m|=I_{\Xi}$,
where $|m\rangle =|m_{1}m_{2}\cdots m_{M}\rangle$ with
$\sigma_{k}^{z}|m_{k}\rangle=(-1)^{m_{k}}|m_{k}\rangle$. Since
$\sum_{k=1}^{M}g_{k}\sigma_{k}^{z}|m\rangle=\sum_{k=1}^{M}(-1)^{m_{k}}g_{k}|m\rangle$,
therefore,
\begin{equation}
H_{\mathcal{T}}=\sum\limits_{m=0}^{2^M-1}(H_{S}+\frac{1}{2}B_{m}\sum\limits_{i=1}^{N}\sigma_{i}^{z})\otimes
|m\rangle \langle m|
\end{equation}
with $B_{m}=\sum\limits_{k=1}^{M}(-1)^{m_{k}}g_{k}$. The time
evolution operator for the combined system-environment is
$U(t)=exp(-itH_{\mathcal{T}})$, i.e.,
\begin{equation}
U(t)=\sum\limits_{m=0}^{2^M-1}U_{m}(t)\otimes |m\rangle \langle m|
\end{equation}
where $U_{m}(t)=exp(-itH_{S}^{(m)})$ with
$H_{S}^{(m)}=H_{S}+\frac{1}{2}B_{m}\sum_{i=1}^{N}\sigma_{i}^{z}$.
We consider the initial state of the spin chain together with the
$M$ independent environment spins of the form as W.H. Zurek has
considered in \cite{Zurek}:
\begin{equation}
|\psi_{S\Xi}(0)\rangle=|\psi_{S}(0)\rangle \otimes
\sum\limits_{m=0}^{2^M-1}c_{m}|m\rangle
\end{equation}

After arbitrary time $t$, the evolution of the spin-environment
system is
$|\psi_{S\Xi}(t)\rangle=U(t)|\psi_{S\Xi}(0)\rangle=\sum_{m=0}^{2^M-1}(U_{m}(t)\otimes
|m\rangle \langle m|)|\psi_{S\Xi}(0)\rangle$. Therefore, the final
density matrix of the target spin is
\begin{eqnarray}
\nonumber \rho'_{N}(t)&=&Tr_{\bar{N}}Tr_{\Xi}(|\psi_{S\Xi}(t)\rangle\langle\psi_{S\Xi}(t)|)\\
\nonumber &=&\sum\limits_{m=0}^{2^M-1}|c_{m}|^2Tr_{\bar{N}}\{
U_{m}(t)|\psi_{S}(0)\rangle\langle\psi_{S}(0)|U^{\dag}_{m}(t)\}\\
&=&\sum\limits_{m=0}^{2^M-1}|c_{m}|^2\rho^{(m)}_{N}(t)
\end{eqnarray}

We note that
$[\frac{1}{2}B_{m}\sum_{i=1}^{N}\sigma_{i}^{z},H_{S}]=0$. It can be
observed that if the the environment is in the state $|m\rangle$,
the output just undergoes a $Z$-rotation by some angle $B_m$, i.e.,
\begin{equation}
\rho^{(m)}_{N}(t)=\left(
\begin{array}{cccc}
1-|f_{1N}(t)|^2 & \alpha f_{1N}^{*}(t)e^{-iB_{m}t}\\
\alpha^{*}f_{1N}(t)e^{iB_{m}t} & |f_{1N}(t)|^2
\end{array} \right)
\end{equation}
Consequently, the only change to the final density matrix caused
by the decoherence is to reduce the off-diagonal elements by a
factor of $\gamma(t)=\sum_{m=1}^{2^{M}-1}|c_m|^2e^{iB_{m}t}$ and
$\gamma^*(t)$. The efficiency of the quantum spin channel is
characterized by the fidelity averaged over all pure state in the
Bloch sphere, that is $F'(t)=\frac{1}{4\pi}\int
Tr[\rho'_{N}(t)|\varphi_{in}\rangle\langle\varphi_{in}|]d\Omega$.

We denote the character function of the environment as
$\eta(B)=\sum\limits_{m=0}^{2^M-1}|c_{m}|^2\delta(B-B_{m})$. After
some straightforward calculation, we can write the average
fidelity of the quantum spin channel in the spin environment as
\begin{equation}
F'(t)=\frac{1}{2}+\frac{|f_{1N}(t)|^2}{6}+\frac{|f_{1N}(t)|}{3}\int
\cos(B t+ \phi)\eta(B) dB
\end{equation}
where $\phi=arg\{f_{1N}(t)\}$. For a general spin environment of
large $M$ values, the character function $\eta(B)$ is
approximately Gaussian \cite{Zurek}, that is
$\eta(B)=exp(-B^2/\vartheta)/\sqrt{\pi\vartheta}$. Then $\int
\cos( B t)\eta(B) dB=e^{-\vartheta t^2/4}$, and the average
fidelity becomes
\begin{equation}
F'(t)=\frac{1}{2}+\frac{|f_{1N}(t)|^2}{6}+\frac{|f_{1N}(t)|\cos{\phi}}{3}e^{-\vartheta
t^2/4}
\end{equation}

\textbf{\textit{Heisenberg spin chain}} We first consider the S.
Bose primitive scheme. There are $N$ spins in the line with
ferromagnetic Heisenberg interactions, labeled $1,2, \cdots, N$.
The Hamiltonian of the spin chain \cite{Bose} is $
H_{S}=-J\sum\limits_{i=1}^{N-1}\vec{\sigma}_{i}\cdot\vec{\sigma}_{i+1}-B\sum\limits_{i=1}^{N}\sigma_{i}^{z}$.
For the situation without considering the influence of decoherence
environment, by choosing the magnetic fields $B$ as some special
value $B_{c}$, one can make the transfer fidelity
$f_{1N}(t)=\langle\textbf{N}|e^{-itH_{S}}|\textbf{1}\rangle\in\mathcal{R}$,
i.e. $\phi=arg\{f_{1N}(t)\}=0$, and then maximized the original
average fidelity. Therefore, the average fidelity of the quantum
spin channel in the spin environment is
\begin{equation}
F'(t)=\frac{1}{2}+\frac{f_{1N}^2(t)}{6}+\frac{f_{1N}(t)}{3}e^{-\vartheta
t^2/4}
\end{equation}

\begin{figure}[htb]
\epsfig{file=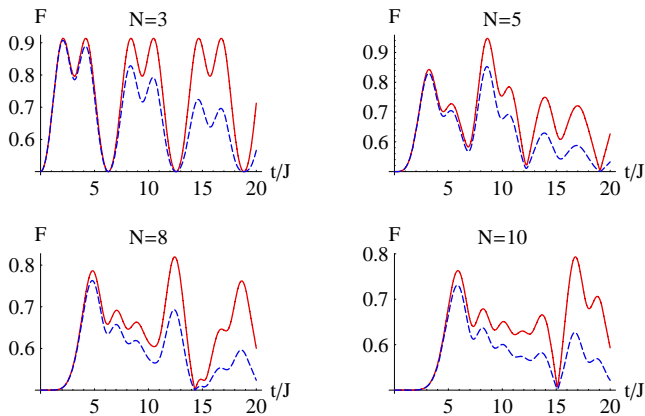,width=8.5cm} \caption{(Color online)
The average fidelity of quantum spin channels in the one common
spin environment (Dashed) and without decoherence (Solid) as
functions of time $t/J$ for different distance $N=3, 5, 8, 10$.
The Gaussian parameter $\vartheta/J^{2}=0.02$. }
\end{figure}

We depict the above average fidelity for $N=3, 5, 8, 10$ in Fig. 2.
Compared with the original fidelity \cite{Bose}
$F(t)=1/2+f_{1N}^2(t)/6+f_{1N}(t)/3$, it can be seen that the
decoherence environment will obviously reduce the efficiency of
quantum communication through the spin chain channels, especially
for the target spin of long distance. Because if the time of
transfer is longer, the decoherence effects are more severe and more
quantum state information will be lost. In fact, without the
influence of decoherence, the critical spin chain length is
$N_{c}=80$, i.e., if the spin chain length $N\leq N_{c}$, the
fidelity of quantum communication through the spin channel will
exceed $2/3$, which is the highest fidelity for classical
transmission of the state \cite{CTF}. However, even the environment
parameter $\vartheta$ is small, the critical spin chain length is
significantly reduced. We list the critical spin chain length for
several environment parameters in the following table. Therefore, if
the spin chain length is large, to achieve satisfactory efficiency
of quantum state transfer, new constraints, e.g. larger coupling
strength $J$, is necessary.

\begin{figure}[htb]
\epsfig{file=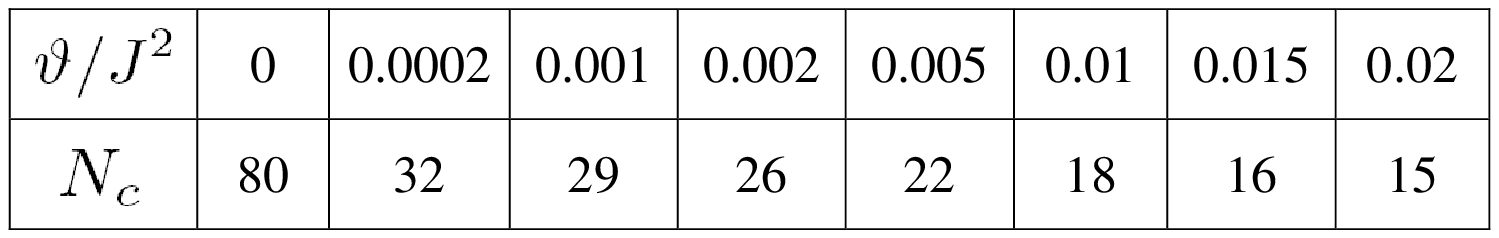,width=8cm} \caption{Critical spin chain
length $N_{c}$ for different environment parameters
$\vartheta/J^2$.}
\end{figure}

We now investigate entanglement distribution through the above
open ended spin channel in the one common spin environment. Two
particles $\mathcal{A}$ and $\mathcal{B}$ are initially in the
entangled state
$|\psi_{\mathcal{A}\mathcal{B}}^{\dag}\rangle=(|01\rangle+|01\rangle)/\sqrt{2}$.
We set $\mathcal{B}$ as the first site of the spin chain channel,
then after some time $t$ entanglement will be established between
$\mathcal{A}$ and the target spin, i.e. the $Nth$ spin. What we
are interested in is the amount of distributed entanglement
between $\mathcal{A}$ and the $Nth$ spin. As pointed out in
\cite{Bose}, the spin chain, without the influence of environment,
acts as an amplitude damping quantum channel, i.e.
\begin{equation}
\rho_{\mathcal{A}N}(t)=\sum\limits_{i=0,1}(I \otimes
M_{i})|\psi_{\mathcal{A}\mathcal{B}}^{\dag}\rangle
\langle\psi_{\mathcal{A}\mathcal{B}}^{\dag}|(I \otimes M_{i}^{\dag})
\end{equation}
with $M_{0}=|0\rangle\langle0|+f_{1N}(t)|1\rangle\langle1|$ and
$M_{1}=(1-|f_{1N}(t)|^2)^{1/2}|0\rangle\langle1|$. In the same way
as discussed above, the final entangled state becomes
$\rho'_{\mathcal{A}N}(t)=\{(1-\lambda^2)|00\rangle\langle00|+\lambda^2|01\rangle\langle01|
+|10\rangle\langle10|+\zeta|01\rangle\langle10|+\zeta^{*}|10\rangle\langle01|\}/2
$, where $\lambda=|f_{1N}(t)|$ and $\zeta=\int \lambda exp(-i B'
t)\eta(B') dB'=\lambda e^{-\vartheta t^2/4}$. Therefore, the
distributed entanglement measured by concurrence \cite{ConC} is,
\begin{equation}
\xi'=\xi_{0} e^{-\vartheta t^2/4}
\end{equation}
where $\xi_{0}=\lambda$ is the distributed entanglement without
decoherence.

\textbf{\textit{Mirror-periodic Hamiltonian}} Now we consider the
perfect state transfer channels, i.e., the mirror-periodic
Hamiltonian scheme in one common spin environment. The $N$ spins
in the line with XY coupling is described by the Hamiltonian
\cite{Christandl}
$H_{S}=\sum_{i=1}^{N-1}\frac{J_{i}}{2}(\sigma^{x}_{i}\sigma^{x}_{i+1}+\sigma^{y}_{i}\sigma^{y}_{i+1})$,
where $J_{i}=\omega\sqrt{i(N-i)}/2$. The most important property
of the mirror-periodic Hamiltonian is that
$e^{-itH}\psi(s_{1},s_{2},\ldots,
s_{N-1},s_{N})=(\pm)\psi(s_{N},s_{N-1},\ldots, s_{2},s_{1})$ for
some time $t$. For the above Hamiltonian $H_{S}$, the transfer
fidelity is $f_{1N}(t)=[-i\sin{(\omega t/2)}]^{N-1}$, i.e. perfect
quantum state transfer will be achieved at a constant time
$t=\pi/\omega$ for arbitrary spin chain distance. According to the
above Eqs.(8, 9),  we can easily get the average fidelity
\begin{equation}
F'(t)=\frac{2}{3}+\frac{1}{3}e^{-\vartheta t^2/4}
\end{equation}
Since the optimal transfer time $t=\pi/\omega$ is constant, the
average fidelity is independent on the spin chain length $N$. This
is different from the situation of Heisenberg spin chain channel.

\section{Local independent environment}
In this section, we will consider another representative decoherence
model, the local independent environment. Each individual spin of
the central system $S$ interacts independently with the local
environment $\Xi_{i}$, as depicted in Fig. 4. The decoherence
process of the multi-spin system can be described by a general
quantum master equation of Lindblad form \cite{ME},
\begin{equation}
\frac{\partial}{\partial
t}\rho=-i[H_{S},\rho]+\sum\limits_{i=1}^{N}
(\mathbf{I}\otimes\cdots\otimes\mathbf{I}\otimes\mathcal{L}_{i}\otimes\mathbf{I}\otimes\cdots\otimes\mathbf{I})\rho
\end{equation}
The superoperator $\mathcal{L}_{i}$ describes the independent
interaction of the $ith$ spin with the local environment.

\begin{figure}[htb]
\epsfig{file=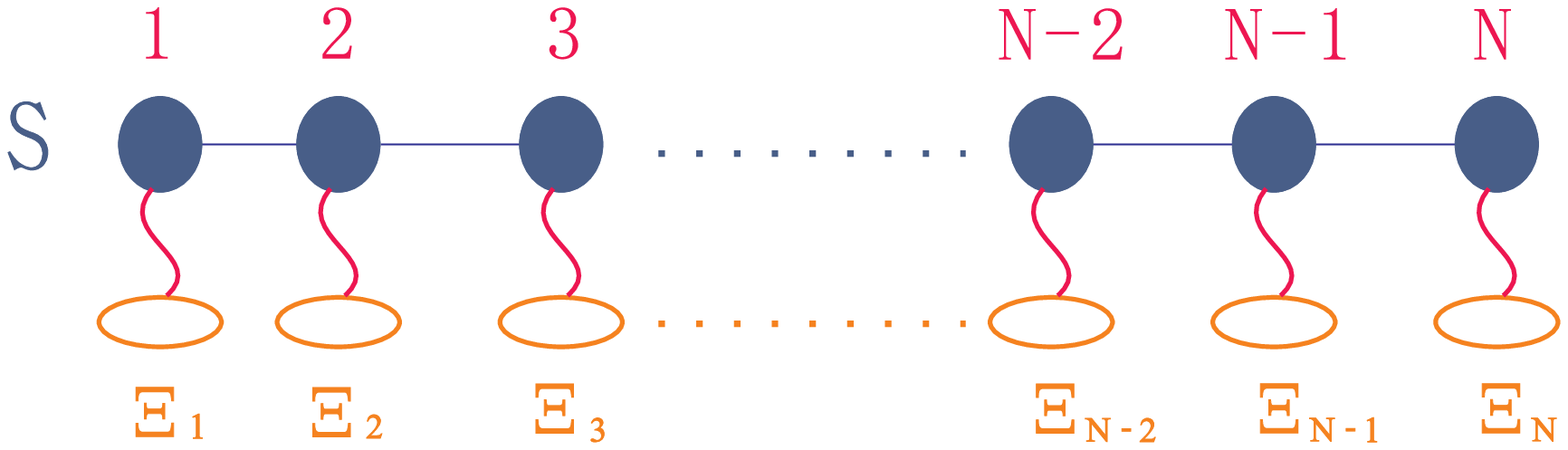,width=8cm} \caption{(Color online)  A quantum
wire with $N$ spins in the line coupled with local independent
environment.}
\end{figure}

It is known that macroscopic systems are more fragile under the
influence of the decoherence environment. In the situation of local
independent decoherence environment, this can be demonstrated in an
explicit way as follows. The phenomenological analysis solution of
the quantum master equation in Eq.(15) can be written as
$\rho(t)=\rho(0)+\int_{0}^{t}\{-i[H_{S},\rho(t')]+\sum_{i=1}^{N}\mathcal{L}_{i}\rho(t')dt'\}$.
And the ideal state without decoherence is
$\rho_{0}(t)=\rho(0)+\int_{0}^{t}\{-i[H_{S},\rho_{0}(t')]\}$. If the
time $t=\delta t$ is short enough, the difference between the real
and ideal state of the central system is
$\rho(t)-\rho_{0}(t)=\sum_{i=1}^{N}\mathcal{L}_{i}\rho(0)\delta t$.
Therefore, it is obvious that the state deviation will become larger
as $N$ increases for most kinds of decoherence model.

\textbf{\textit{Mirror-periodic Hamiltonian}} We first consider the
mirror-periodic Hamiltonian scheme in the local independent
dephasing and damping channels.

\textit{Dephasing Channel} The dephasing process corresponds to the
situation where only phase information is lost \cite{Chuang},
without energy exchange. The superoperator for the dephasing channel
\cite{IDE} is
\begin{equation}
\mathcal{L}_{i}\rho=-\frac{\gamma_{i}}{2}(\rho-\sigma_{i}^{z}\rho\sigma_{i}^{z})
\end{equation}
For simplicity, we assume the system-environment coupling strength
$\gamma_{i}=\gamma$ are the same for all spins.

In the case of quantum state transfer, the initial state of the
system is
$|\psi_{S}(0)\rangle=\alpha|\mathbf{0}\rangle+\beta|\mathbf{1}\rangle$.
The state transfer dynamics under the influence of dephasing
channel is completely determined by the evolution in the
\textit{zero} and \textit{single} excitation subspace
$\mathcal{H}_{\mathrm{0}\oplus \mathrm{1}}$. Therefore, we only
need to solve the above master equation in this
$(N+1)-$dimensional subspace. When restricted to the subspace
$\mathcal{H}_{\mathrm{0} \oplus \mathrm{1}}$,
\begin{eqnarray}
\nonumber
H_{S}&=&\sum\limits_{i=1}^{N-1}J_{i}(|\mathbf{i}\rangle\langle\mathbf{i+1}|+|\mathbf{i+1}\rangle\langle\mathbf{i}|)\\
\sigma^{z}_{i}&=&\mathbf{I}_{\mathrm{N+1}}-2|\mathbf{i}\rangle\langle\mathbf{i}|
\end{eqnarray}

At time $T=\pi/\omega$, the reduced density of the $Nth$ spin is
\begin{equation}
\rho^{(N)}(\pi/\omega)=\left(
\begin{array}{cccc}
1-\rho_{NN}(\pi/\omega) & \rho_{0N}(\pi/\omega)\\
\rho^{*}_{0N}(\pi/\omega) & \rho_{NN}(\pi/\omega)
\end{array} \right)
\end{equation}
where
$\rho_{NN}(\pi/\omega)=\langle\mathbf{N}|\rho(\pi/\omega)|\mathbf{N}\rangle$
and
$\rho_{0N}(\pi/\omega)=\langle\mathbf{0}|\rho(\pi/\omega)|\mathbf{N}\rangle$.
Therefore, the probability of an excitation transfer from the $1st$
spin to the $Nth$ spin at time $t=\pi/\omega$ is
$P=\rho_{NN}(\pi/\omega)$. And the fidelity between the real and
ideal transferred quantum state, i.e.
$|\psi_{ideal}(\pi/\omega)\rangle=\alpha|0\rangle+(-i)^{N-1}\beta|1\rangle$,
is
\begin{eqnarray}
\nonumber
F(\pi/\omega,\alpha,\beta)&=&(1-\rho_{NN}(\pi/\omega))|\alpha|^2+\rho_{NN}(\pi/\omega)|\beta|^2\\
&+&2Re\{(-i)^{N-1}\rho_{0N}(\pi/\omega)\alpha^{*}\beta\}
\end{eqnarray}

\begin{figure}[htb]
\epsfig{file=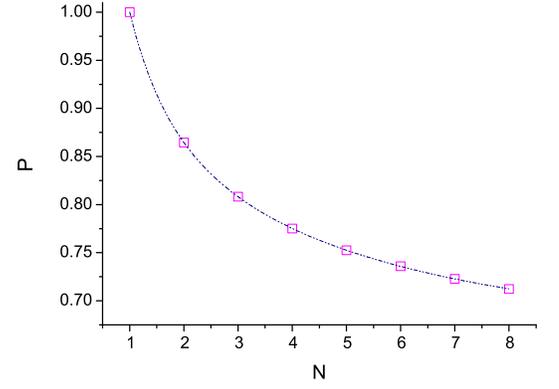,width=7cm} \caption{(Color online) The
probability of an excitation transfer as function of the spin chain
length $N$ for the dephasing channel in the local independent model.
The system-environment coupling strength $\gamma/\omega=0.1$.}
\end{figure}

\begin{figure}[htb]
\epsfig{file=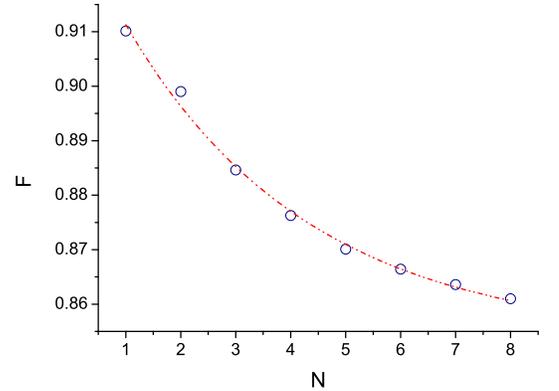,width=7cm} \caption{(Color online) The
fidelity as function of the spin chain length $N$ for the dephasing
channel in the local independent model. The system-environment
coupling strength $\gamma/\omega=0.1$.}
\end{figure}

The efficiency of quantum communication through the above quantum
spin channel is characterized by the average fidelity over all pure
states in the Bloch sphere $F(\pi/\omega)=\frac{1}{4\pi}\int
F(\pi/\omega,\alpha,\beta) d\Omega$. In the mirror-periodic
Hamiltonian scheme, we solve the above differential equations
numerically, and depict the probability of an excitation transfer
and the average fidelity in Fig. 5 and 6. Though the transfer time
is constant $t=\pi/\omega$ for any spin chain distance $N$, the
probability of an excitation transfer and the average fidelity will
still decay as the spin chain length increases. Due to the
decoherence effects, the mirror-periodic Hamiltonian scheme can not
achieve perfect quantum state transfer again. However, using larger
nearest-neighbor interaction $\omega$, i.e. relative smaller
system-environment coupling strength $\gamma/\omega$, we will
increase the efficiency of quantum communication in the decoherence
environment.

\textit{Damping channel} The pure damping channel corresponds to
the decay process of the central system coupled to a thermal bath
at zero temperature \cite{Chuang,IDE}. The superoperator for the
damping channel is
\begin{equation}
\mathcal{L}_{i}\rho=-\frac{\gamma_{i}}{2}(\sigma^{-}_{i}\sigma^{+}_{i}\rho+\rho\sigma^{-}_{i}\sigma^{+}_{i}
-2\sigma_{i}^{+}\rho\sigma_{i}^{-})
\end{equation}
where $\sigma^{\pm}_{i}=(\sigma^{x}_{i}\pm i\sigma^{y}_{i})/2$,
with the system-environment coupling strength $\gamma_{i}=\gamma$
for all spins.

In the similar way as the dephasing channel, When restricted to
the subspace $\mathcal{H}_{\mathrm{0} \oplus \mathrm{1}}$,
\begin{equation}
\sigma^{-}_{i}=|\mathbf{i}\rangle\langle\mathbf{0}|,
\sigma^{+}_{i}=|\mathbf{0}\rangle\langle\mathbf{i}|,
\sigma^{-}_{i}\sigma^{+}_{i}=|\mathbf{i}\rangle\langle\mathbf{i}|
\end{equation}
The reduced density of the $Nth$ spin and the average fidelity in
Eqs.(18, 19) are applicable to the damping channel too. By solving
the corresponding differential master equations numerically, we can
obtain the probability of an excitation transfer and the average
fidelity for different spin chain length $N$. Unlike the situation
of the dephasing channel, we find that at time $t=\pi/\omega$, the
probability of an excitation transfer is independent on the spin
chain length, that is $\rho_{NN}(\pi/\omega)=e^{-\gamma\pi}$.
Moreover, the average fidelity $F(\pi/\omega)$ is independent on the
spin chain length either. This result is somewhat surprising.
Whether this property is only hold by the mirror-periodic
Hamiltonian in a pure damping environment is an interesting open
problem, which will be investigated in details in our following
work.

\textbf{\textit{Heisenberg spin chain}} For the Heisenberg spin
chain channel in the local independent dephasing and damping
environment, we can express the system Hamiltonian $H_{S}$ in the
subspace $\mathcal{H}_{\mathrm{0} \oplus \mathrm{1}}$ and solve
the quantum master equations in the similar way as discussed
above. We list the maximum probability of an excitation transfer
$P$ for different spin chain length $N$ in the following table.
The behavior of the average fidelity $F$ is similar to the
probability of an excitation transfer $P$.

\begin{figure}[htb]
\epsfig{file=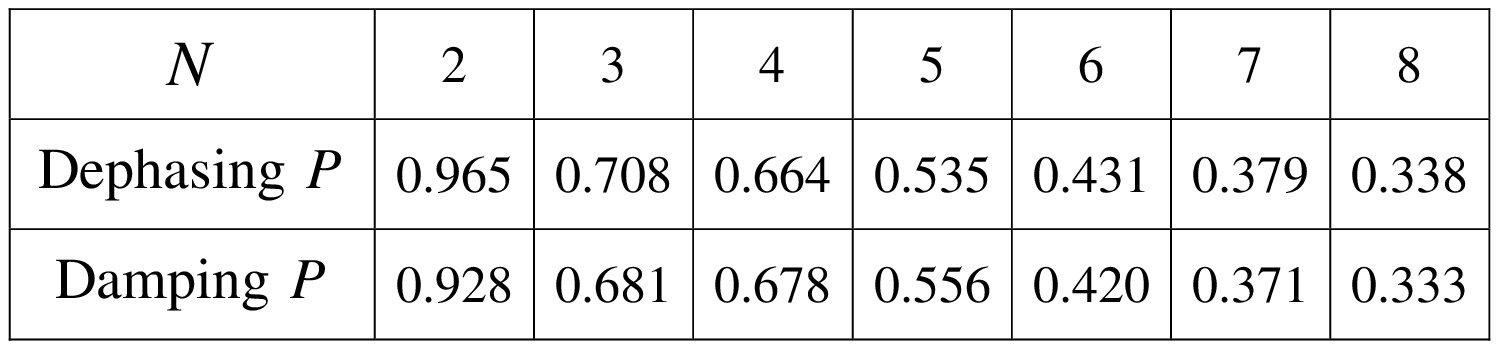,width=8cm} \caption{The maximum
probability of an excitation transfer for different spin chain
length $N$ in the local independent model. The system-environment
coupling strength $\gamma /J=0.1$.}
\end{figure}

It can be seen that, the maximum probability of an excitation
transfer, i.e. the efficiency of the Heisenberg spin chain
channel, will decay as the spin chain length $N$ increases not
only for the dephasing channel but also for the damping channel.
This is slightly different from the situation of mirror-periodic
Hamiltonian, where the probability of an excitation transfer and
the average fidelity are independent on the spin chain length.

\section{Conclusions and discussions}
In conclusion, we have investigated the efficiency of quantum
communication through the spin chain channels under the influence of
a decoherence environment. We focus on the Heisenberg spin chain and
the mirror-periodic Hamiltonian scheme. Two representative
decoherence models are considered, one is the one common spin
environment, the other is local independent environment. It can be
seen that the efficiency of the quantum wires will be significantly
lowered by the external decoherence environment. Generally speaking,
the decoherence effects become more serious for larger spin chain
length. However, for different spin chain Hamiltonian and
decoherence models, the results will be somewhat different.

In Refs \cite{Daniel}, it has been shown that in some specific
environment, quantum state transfer is possible with the same
fidelity and only reasonable slowing. However, for more general
decoherence models as considered in this paper, we show that, to
achieve highly efficient long distance quantum state transfer in a
decoherence environment, the time of transfer \cite{Time} becomes a
crucial factor and more constraints on the spin system Hamiltonian
are requisite. We should resort to new encoding strategy for the
protection of quantum state information during the transfer along
the spin chain channels. Besides, we provide some possible evidence
for the unusual dynamical decoherence properties of the
mirror-periodic Hamiltonian, which may lead to some valuable
utilities of this special class of Hamiltonian.

\section{Acknowledgments}
We thank Dr. Yong Hu and Daniel Burgarth for helpful discussions and
valuable suggestions. This work was funded by National Fundamental
Research Program (2001CB309300), the Innovation funds from Chinese
Academy of Sciences, NCET-04-0587, and National Natural Science
Foundation of China (Grant No. 60121503, 10574126).

\end{document}